\documentclass[twocolumn,preprintnumbers,amsmath,amssymb]{revtex4}

\usepackage{graphicx}
\usepackage{dcolumn}
\usepackage{bm}

\begin{document}

\title{Regge Trajectories of Quark Gluon Bags}
\author{K. A. Bugaev and G. M. Zinovjev}%
\affiliation{Bogolyubov Institute for Theoretical Physics, 
Kiev, Ukraine}

\email{bugaev@th.physik.uni-frankfurt.de; Gennady.Zinovjev@cern.ch}

\setcounter{page}{1}%

%
%

\def\ts{t{\ss}s}
\def\ss{\hspace{.5pt}}
%
%
%
%

\begin{abstract}
Using an exactly solvable statistical model   we discuss  the equation of state of large/heavy and short-living   quark gluon plasma (QGP) bags.
We argue   that the  large width of the QGP bags  explains   not only
the observed deficit in the number of  hadronic resonances, but also clarifies the reason   why
the heavy QGP bags   cannot be directly observed  even  as metastable  states in a hadronic phase.
Also the Regge trajectories of  large and heavy  QGP bags  are established both in a vacuum and in a strongly interacting medium.
It is  shown  that    at  high  temperatures   the average mass and width of the QGP bags  behave in accordance with   the upper bound of the Regge trajectory asymptotics (the linear asymptotics), whereas  for  temperatures  below $T_H/2$  ($T_H$ is the Hagedorn temperature)  they obey  the lower bound  of   the Regge trajectory asymptotics (the square root one).
Thus, for   $T < T_H/2$ the spin of the QGP bags is restricted from above, whereas
for    $T> T_H/2$  these  bags demonstrate the standard  Regge behavior consistent with the string  models.

\hfill \\

\end{abstract}


\setlength{\topmargin}{-1.0cm}

\maketitle


\section{Introduction}

Regge poles have been introduced in particle physics  before the  QCD era and since  the beginning of 60-ies  \cite{Reg:1} they  are widely used to  describe  the high-energy
interactions of hadrons and nuclei.   Regge approach establishes an important
connection between high energy scattering and spectrum of particles and resonances. It served as a starting point  to  introduce  the  dual and string models of  hadrons.
Up to now a rigorous derivation of Regge poles in QCD remains  an unsolved  problem since it is  related to the nonperturbative effects in QCD and the problem of confinement.

Nowadays  the Regge trajectories  are widely understood as the linear relation between the resonance mass square  and resonance  spin or radial quantum number, whereas  the Regge trajectory $\alpha (S_r) $  contains information about the resonance mass  $M_r$ and width $\Gamma_r$.  Indeed,  the   resonance spin $J$  is  defined in the complex energy plane as $J = \alpha \left( (M_r  -  \frac{i}{2}\Gamma_r)^2 \right)  $.
Moreover,  the linear trajectories, i.e. $\alpha (S_r)  \sim S_r$, which follow from the  string models, are often believed to be the only Regge trajectories of hadrons.

However,  long ago it was shown that under the plausible assumptions the linear Regge trajectories correspond   to  the upper bound of the asymptotic behavior, whereas  its lower bound  is
given by the square root trajectory, i. e. $\alpha_l (S_r) \sim [- S_r]^{\frac{1}{2}} $
\cite{Khuri, Trushevsky:77}.
Moreover, there were some indications \cite{Trushevsky:77}  that the square root trajectory
should give the asymptotic behavior of excited hadronic resonances.
The latter means that for each family of hadronic  resonances  the Regge poles do not go beyond  some vertical line in the complex spin plane, i.e.
 in asymptotics $S \rightarrow + \infty$ the resonances should become infinitely wide.

Since the linear Regge trajectories of hadrons generate  the Hagedorn mass spectrum \cite{Miranski:73}, the square root ones should  lead to a weaker growth of hadronic mass spectrum.
At first glance  it seems that the  experimental mass spectrum of hadrons  \cite{Bron:04}
does not show an exponential increase  at hadron masses above 2.5 GeV and, hence,
it evidences against the linear Regge trajectories of heavy hadrons.
Moreover, the best  description of particle yields observed in a very wide range of
collision  energies of heavy ions   is  achieved
by the statistical model which incorporates  all hadronic resonances not heavier than 2.3 GeV \cite{HG}.
Again  it looks like that  heavier hadronic species, except for the long living ones, are simply absent in  the experiments \cite{Blaschke:03}.
Thus, we are confronted with a serious  conceptual  problem between a few  theoretical expectations   and  several   experimental facts.

Recently this conceptual problem was resolved within the finite width model (FWM)
\cite{FWM:08}.  The  FWM  introduces the  medium dependent finite width of QGP  bags
into  an exactly solvable statistical  model.
It  shows  that the  large width of the QGP bags   explains  not only
the observed deficit in the number of  hadronic resonances, but also clarifies the reason   why  the heavy QGP bags  and strangelets  cannot be directly observed  even as metastable  states in a hadronic phase.
Also the FWM allows one to establish  \cite{Reggeons:08} the  Regge trajectories of  large and heavy  QGP bags   both in a vacuum and in a strongly interacting medium.
As will be shown below at  high  temperatures the average mass and width of the QGP bags  behave in accordance with   the upper bound of the Regge trajectory asymptotics (the linear asymptotics), whereas at low temperatures   they obey  the lower bound  of   the Regge trajectory asymptotics (the square root one).
Thus, for  low  temperatures  the spin of the QGP bags is restricted from above, whereas
for  high  temperatures    these  bags demonstrate the typical Regge behavior consistent with the string  models.

The work is organized  as follows.  Sec. 2  discusses   the main ideas and  results of FWM.
The Regge trajectories of QGP bags are established in Sec. 3, while our conclusions are given in Sec. 4.

\section{FWM of QGP Bags}

The main object of the FWM is the mass-volume spectrum of heavy and large QGP bags
at temperature $T$
($V_0 \approx 1$ fm$^3$, $M_0 \approx 2.5$ GeV \cite{FWM:08,Reggeons:08})
\begin{equation}  \label{FsHQ}
F_Q(s,T) = \int\limits_{V_0}^{\infty}dv\hspace*{-0.1cm}\int\limits_{M_0}^{\infty}
 \hspace*{-0.1cm}dm~\rho(m,v)\exp(-sv)\phi(T,m)~,
\end{equation}
where  the thermal  density of bags  of mass $m$   reads as
\begin{equation}  \label{phi}
\phi(T,m) = \int\limits_0^{\infty}\hspace*{-0.0cm}p^2dp~
\exp{\textstyle \left[- \frac{(p^2~+~m^2)^{1/2}}{T} \right] } \,.
\end{equation}
In (\ref{FsHQ})   $s$ denotes  the  variable of the isobar ensemble which is dual to the system volume.
An exponential $\exp(-sv)$ in (\ref{FsHQ}) describes the hard core repulsion between the bags \cite{FWM:08}, while
the density of states $\rho(m,v)$  of bags of mass $m$ and volume $v$ has the form
\begin{eqnarray}\label{Rfwm}
 \rho (m,v)  & = &  \frac{ \rho_1 (v)  ~N_{\Gamma}}{\Gamma (v) ~m^{a+\frac{3}{2} } }
 \exp{ \textstyle \left[ \frac{m}{T_H}   -   \frac{(m- B v)^2}{2 \Gamma^2 (v)}  \right]  } \,, \\
  \rho_1 (v)  & = & f (T)\, v^{-b}~ \exp{\textstyle \left[  -  \frac{\sigma(T)}{T} \, v^{\varkappa}\right] }\,.
\label{R1fwm}
 \end{eqnarray}
As one can see from (\ref{Rfwm}) the the density of states  has a Hagedorn like parameterization  with respect to mass   and  the Gaussian attenuation  around the bag mass
$B v$ ($B$ is the mass density of a bag of a  vanishing width) with the volume dependent  Gaussian  width
$\Gamma (v)$ or width hereafter.
We will distinguish it from the true width defined as
$\Gamma_R = \alpha \, \Gamma (v)$ ($\alpha \equiv 2 \sqrt{2 \ln 2}\,$).
As shown in  \cite{FWM:08, Reggeons:08}
the Breit-Wigner attenuation  of  a resonance mass cannot be used in the spectrum
(\ref{Rfwm}) because in this  case the  finite width  leads to a divergency of the mass integral in (\ref{FsHQ}) above  $T_H$.

The normalization factor
obeys the condition

\vspace*{-0.5cm}
\begin{eqnarray}\label{Ng}
& N_{\Gamma}^{-1}~ = ~ \int\limits_{M_0}^{\infty}
 \hspace*{-0.1cm} \frac{dm}{\Gamma(v)}
    \exp{\textstyle \left[  -   \frac{(m- B v)^2}{2 \Gamma^2 (v)}  \right] } \,.
 \end{eqnarray}
The constants  $a > 0$ and  $b > 0$ are discussed in \cite{FWM:08}.

Also the volume spectrum in  (\ref{R1fwm}) contains the surface free energy (${\varkappa} = 2/3$) with the $T$-dependent
surface tension which can be  parameterized   in a general  way \cite{Bugaev:04b, Bugaev:07, CritPoint:09}
\begin{eqnarray}\label{Sigma}
\sigma(T, \mu) =
\left\{ \begin{array}{rr}
\sigma^- > 0  \,,  &\hspace*{0.1cm}
T \rightarrow T_{\Sigma} (\mu)  - 0 \,,\\
 0 \,, &\hspace*{0.1cm}  T =  T_{\Sigma} (\mu) \,, \\
\sigma^+ < 0 \,,  &\hspace*{0.1cm}
T \rightarrow T_{\Sigma} (\mu)  + 0  \,.
\end{array} \right.
\end{eqnarray}
For $T \le T_{\Sigma} (\mu)$  such a parameterization  is justified by the usual  cluster models
like the FDM \cite{Fisher:67} and SMM \cite{Bondorf:95,Bugaev:00,Bugaev:01,Reuter:01}, whereas
the general consideration  for any  $T$   can be driven  by  the surface partitions of the Hills and Dales model
\cite{Bugaev:04b}.

The recent results obtained within the exactly solvable models \cite{Bugaev:07,CritPoint:09}
also justify  the parameterization  (\ref{Sigma}) and show that
the only physical reason of degeneration of the first order deconfinement phase transition
 at low baryonic densities  into a cross-over
is negative surface tension coefficient in this region.
Moreover,  an existence of negative surface tension  at the cross-over region
was  directly demonstrated from the lattice QCD data  very recently  within a new phenomenological model of  the confinement   \cite{SurfTension:09}.

An actual choice of the continuous functions $\sigma^\pm $ of temperature  $T$  and baryonic chemical potential  $ \mu$  along with parameterization of the nil line of the surface tension coefficient  $T_{\Sigma} (\mu)$ for the tricritical  and critical endpoint of
the QCD phase diagram can be found in \cite{Bugaev:07} and \cite{CritPoint:09}, respectively.

The spectrum (\ref{Rfwm}) has a simple form, but is rather general since both the width $\Gamma (v)$ and the bag mass density $B$ can be medium dependent. It clearly reflects the fact
that the QGP bags are similar to   the ordinary  quasiparticles with the medium dependent characteristics (life-time, most probable values of  mass and volume).
In principle, one could consider various $v$-dependences $\Gamma (v)$, but
in  \cite{FWM:08, Reggeons:08}  it was shown that the only  square root dependence of the resonance width on its   volume, i.e.  $\Gamma(v)  = \gamma v^\frac{1}{2}$, does not lead to the problems with the large  QGP bags existence.

For large bag volumes ($v \gg M_0/B > 0$) the factor (\ref{Ng})  can be
found as  $N_\Gamma \approx 1/\sqrt{2 \pi} $.   Similarly, one can show that  for heavy free bags  ($m \gg B V_0$, $V_0 \approx 1$ fm$^3$ \cite{FWM:08},
ignoring the  hard core repulsion and thermostat)
\vspace*{-0.05cm}
\begin{eqnarray}\label{Rm}
& \rho(m)  ~ \equiv   \int\limits_{V_0}^{\infty}\hspace*{-0.1cm} dv\,\rho(m,v) ~\approx ~
\frac{  \rho_1 (\frac{m}{B}) }{B ~m^{a+\frac{3}{2} } }
\exp{ \textstyle \left[ \frac{m}{T_H}     \right]  } \,.
\end{eqnarray}

\vspace*{-0.1cm}
\noindent
It originates in   the fact that  for heavy bags the
Gaussian  in (\ref{Rfwm}) acts like a Dirac $\delta$-function for
$\Gamma(v)  = \gamma v^\frac{1}{2}$.
Similarly to (\ref{Rm}), one can estimate the width of heavy free bags  averaged over bag volumes and get
\begin{eqnarray}\label{WidthFreeBag}
\overline{\Gamma(v) } \approx  \Gamma(m/B) = \gamma
\sqrt{ \frac{m}{B} }\,.
\end{eqnarray}
Thus,
 the mass spectrum of heavy free QGP bags
is  the Hagedorn-like one with the property that  heavy resonances must  have   the large  mean width  and, hence,  they
are hardly   observable.
This resolves the conceptual problem of the deficit of observed  heavy hadronic resonances  compared to the Hagedorn mass  spectrum.
Applying these arguments to the strangelets,
we conclude  that, if their mean volume is a few cubic fermis or larger, they  should survive  a  very short time,
which  is similar to the results of  Ref. \cite{Strangelets:06}  predicting  an instability of such strangelets.

An analysis of  the  full spectrum  (\ref{FsHQ}) allows one to determine the pressure  of the QGP bags
\cite{FWM:08} which shows two drastically different regimes depending the value of the most probable
mass of a bag  ($v \gg V_0$)
\begin{eqnarray}\label{Mprob}
& \langle m \rangle ~ \equiv ~  B v + \Gamma^2 (v) \beta\,,\quad {\rm with}
\quad \beta \equiv  T_H^{-1} - T^{-1} \,.
\end{eqnarray}
For
temperatures below (above)  $T_\pm = c_\pm \,T_H$ ($ 0 < c_\pm < 1$)   the most probable mass is negative (positive) and  the spectrum  (\ref{FsHQ})   defines the low (high) temperature pressure $p^-$ ($p^+$)
\begin{eqnarray}\label{ptot}
p  =
\left\{ \begin{array}{rl}
p^+ \equiv T \left[ \beta B + \frac{\gamma^2}{2 } \beta^2 \right]    \,,  &\hspace*{0.1cm}
\langle m \rangle > 0  \,,\\
p^\pm \equiv    \frac{B T \beta} {2 } \,,   &\hspace*{0.1cm}  \langle m \rangle = 0 \,,   \\
p^-    \equiv   - T\frac{ B^2}{2\, \gamma^2 }  \,,  &\hspace*{0.1cm}
\langle m \rangle  <  0  \,.
\end{array} \right.
\end{eqnarray}
There are two remarkable facts concerning the low temperature pressure.
First, for  $\langle m \rangle  \le  0 $     the  resulting mass attenuation  of the integrand in  (\ref{FsHQ})
decreases at  fixed  bag volume so rapidly that  the only vicinity of $M_0$ contributes into the mass-volume spectrum. In other words, in this regime all heavy  QGP bags   are extremely suppressed  and as a result  only the  smallest bags  with the mass $M_0$ and
the width about $\Gamma (V_0)$  can contribute into the spectrum  (\ref{FsHQ}) \cite{FWM:08}.
Consequently, such QGP bags would not be distinguishable
from the usual low-mass hadrons.
Such a regime leads to the {\it subthreshold suppression of the QGP bags} at low temperatures even in finite systems and, hence, is able to explain the absence of heavy/large QGP bags and strangelets
for $T < T_\pm$.

Second,  for the non-vanishing  functions $\gamma$ and $B$  at  low temperatures, i.e.  $\gamma_0= \gamma(T=0)>0$ and $B_0 =B(T=0) >0 $,  the QGP pressure at such temperatures should be negative and  linear in temperature  $p^- (T \rightarrow 0) \approx - T \frac{ B_0^2}{2\, \gamma_0^2 } $.
Such a  linear $T$-behavior of the QGP pressure at low temperatures,  $p_{QGP} = \sigma_p T^4 - A_1 T $,   is known  for a long time \cite{LQCD:0}  and was reported by several groups (see \cite{Reggeons:08} for details).
Using this fact and matching $p^+$ with $p_{QGP}$  it was possible to estimate the  resonance width coefficient $\gamma$ and
the mass density $B$  from the lattice QCD data \cite{Reggeons:08}
\begin{eqnarray}\label{gammaI}
\gamma^2 (T) & = & 2 \, \beta^{-1} [ \sigma_p T_H T (T^2 + TT_H + T_H^2) - B(T) ]  \,,\\
\label{BI}
B(T) & = &  \sigma_p T_H^2  (T^2 + TT_H + T_H^2)  \,,
\end{eqnarray}
where $3  \sigma_p$ is the Stefan-Bolzmann constant of the QGP.   Eqs. (\ref{gammaI}) and (\ref{BI}) allow one to determine $T_\pm = 0.5 T_H$ and show that  the resonance width  at zero temperature  very weakly depends on the number of elementary degrees of freedom in QGP, but strongly depends on the cross-over temperature $T_{co}$
\begin{equation}
\Gamma_R (V_0, T=0) \approx  C_\gamma \, V_0^{\frac{1}{2}} \, T_{co}^{\frac{5}{2}} \, \alpha \,,
\end{equation}
where the constant $C_\gamma $, depending on number of color and flavor states of QGP,   varies  between $1.22$ and $1.3$ \cite{Reggeons:08}.
The minimal width of the QGP bags strongly increases with temperature. For instance,
at Hagedorn temperature one obtains $\Gamma_R (V_0, T=T_H)   = \sqrt{12}\,\Gamma_R (V_0, T=0)$.
Therefore, for $T_{co} \in [170; 200]$ MeV the  minimal width of the QGP bags is
$\Gamma_R (V_0, T=0) \in [ 400; 600]$ MeV and  $\Gamma_R (V_0, T=T_H)  \in [1400; 2000]$ MeV. These estimates clearly show us that, even without the subthreshold suppression, the heavy/large QGP bags cannot be directly  observed at  any temperature
due to very short life-time.

Also one of  the most  remarkable features of the FWM is  that  its  mass dependence of  a mean resonance width can help  to  determine the Regge trajectories of QGP bags and to resolve a few  problems  related to them.


\section{Asymptotic  Trajectories of QGP Bags}

Nowadays there is  great   interest in  the behavior of the  Regge  trajectories of higher resonances in  the  context  of the
5-dimensional {string   theory holographically dual to QCD}  \cite{AdS} which is known as
anti-de-Sitter space/conformal field theory (AdS/CFT). However, as was mentioned earlier,  the Regge trajectories  are widely understood only  as the linear relation between the resonance mass square  and resonance  spin or radial quantum number.
We  would like to determine  the full Regge trajectories of QGP bags.
In our  analysis  we follow
Ref.
\cite{Trushevsky:77} which is based on the following most general assumptions:
(I) $\alpha(S)$ is an analytical function, having only the physical cut  from
$S= S_0$ to $S = \infty$; (II)  $\alpha(S)$ is polynomially restricted at the whole physical sheet;  (III) there exists a finite limit of the phase trajectory at
$S \rightarrow  \infty$.
Using these assumptions, it was possible to prove \cite{Trushevsky:77}  that  for
$S \rightarrow  \infty$ the upper bound  of  the Regge  trajectory asymptotics
at the whole physical sheet
is
\begin{eqnarray}\label{aUP}
\alpha_u (S) = - g^2_u \left[  - S   \right]^\nu \,,\quad {\rm with} \quad   \nu \le 1 \,,
\end{eqnarray}
where the function  $g^2_u > 0$ should increase slower than any power
in this limit  and   its  phase must  vanish at  $|S| \rightarrow ~\infty$.

On the other hand, in Ref. \cite{Trushevsky:77}  it was also shown that, if
in addition to (I)-(III)
one requires  that the   transition amplitude $T(s,t)$ is
a  polynomially  restricted function of $S$ for all nonphysical $t > t_0 > 0$,
then the real part of the Regge trajectory  does not  increase  at
$|S| \rightarrow ~\infty$ and  the trajectory behaves as
\begin{eqnarray}\label{aLOW}
\alpha_l (S) =  g^2_l \left[ - \left[  - S   \right]^{\frac{1}{2}} + C_l \right] \,,
\end{eqnarray}
where  $g^2_l > 0$ and $C_l$ are some constants.  Moreover,  (\ref{aLOW}) defines the lower bound for the asymptotic behavior of the Regge trajectory \cite{Trushevsky:77}. The  expression (\ref{aLOW})   is a generalization of a well known  Khuri  result \cite{Khuri}. It means that for each family of hadronic  resonances  the Regge poles do not go beyond  some vertical line in the complex spin plane.
In other words,  it means that in asymptotics $S \rightarrow + \infty$ the resonances become infinitely wide, i.e. they are moving out of the real axis of the proper angular momentum $J$ and, therefore,   there is only  a finite number of resonances in the corresponding transition amplitude.

To compare the FWM results with the trajectories  (\ref{aUP}) and (\ref{aLOW})
we need to relate  the  mass and width of QGP bags, since
they  are independent variables in this model. Nevertheless, this can be done for  their  averaged values.

To illustrate this statement,
let's   recall our result  on the mean Gaussian  width of the free bags averaged with respect to their volume (\ref{WidthFreeBag})
by the spectrum (\ref{Rm}).
Using the formalism of  \cite{Trushevsky:77}, it  can  be shown that
at zero temperature
the free QGP bags of mass $m$ and mean resonance  width $\alpha\, \overline{\Gamma(v) } |_{T=0} \approx \alpha\,\gamma_0
\sqrt{ \frac{m}{B_0} }$
precisely correspond  to  the following Regge trajectory
\begin{eqnarray}\label{alphaHT}
&  \hspace*{-0.1cm}
\alpha_r  (S) =  g^2_r [S + a_r (- S)^\frac{3}{4} ] \quad  {\rm with} \quad {a_r =const  < 0}\,.
\end{eqnarray}
Indeed, substituting $S = |S| e^{i \, \phi_r}  $ into (\ref{alphaHT}), then  expanding  the second term on the right hand side of   (\ref{alphaHT}) and requiring ${\rm Im} \left[ \alpha_r (S) \right] = 0$, one finds the phase of
physical trajectory (one of four roots of one fourth power in  (\ref{alphaHT}))
\begin{eqnarray}\label{phiS}
&
\phi_r  (S) \rightarrow  \frac{a_r \sin \frac{3}{4}\pi  }{|S|^\frac{1}{4} }  \rightarrow
0^-\,,
\end{eqnarray}
which is vanishing in the correct quadrant of the complex $S$-plane.
Considering the complex energy plane $E = \sqrt{S} \equiv M_r - i \frac{\Gamma_r}{2}$,
one can  determine the mass  $M_r$ and the width $\Gamma_r$
\begin{eqnarray}\label{mgamI}
M_r \approx |S|^\frac{1}{2} \,, \quad  \Gamma_r \approx - |S|^\frac{1}{2}
\phi_r  (S)   =  \frac{|a_r| M_r^\frac{1}{2}   }{  \sqrt{2} },
\end{eqnarray}
of a resonance belonging to the trajectory  (\ref{alphaHT}).

Comparing the mass dependence of the width in (\ref{mgamI}) with the mean width of  free QGP bags
(\ref{WidthFreeBag}) taken at $T=0$, it is natural  to identify them
\begin{eqnarray}\label{aIpfree}
&
a_r^{free} \approx - \alpha \,\gamma_0 \, \sqrt{\frac{2}{B_0}} = - 4  \, \gamma_0 \, \sqrt{\frac{\ln 2}{B_0}}  \,,
\end{eqnarray}
and to deduce  that   the free QGP bags belong to the linear  Regge trajectory (\ref{alphaHT}).
Such a conclusion is supported and justified by  the well established results on the linear  Regge trajectories of hadronic resonances \cite{RegBook} and
by  theoretical expectations  of  the dual resonance model \cite{DualRM}, the open string model
\cite{Shuryak:sQGP}, the closed string model \cite{Shuryak:sQGP} and  the   AdS/CFT \cite{AdS}.
Moreover,  the most direct way to connect  the FWM bags with  the string models is provided by the recently suggested model  of the confinement phenomenon \cite{SurfTension:09}
which allows us to relate the string tension of confining color tube  and the surface tension of QGP bags.

Next we consider the second  way of averaging the mass-volume spectrum
with respect to  the resonance  mass
\begin{eqnarray}\label{mmeanI}
& 
\overline{m} (v)  ~ \equiv ~  \frac{   \int\limits_{M_0}^{\infty}\hspace*{-0.0cm} dm
\int  \frac{d^3k}{(2\pi)^3}  \,\rho(m,v) ~ m ~ e^{- \frac{\sqrt{k^2 + m^2} }{T}}
}{
 \int\limits_{M_0}^{\infty}\hspace*{-0.cm} dm
\int  \frac{d^3k}{(2\pi)^3}  \,\rho(m,v) ~e^{- \frac{\sqrt{k^2 + m^2} }{T}}
 }
\, ,
\end{eqnarray}
which is technically simpler than averaging with respect to the resonance volume,
but we will make  the necessary  comments  on the other way of  averaging  in the  appropriate places.

Using the results of  the preceding section  one can find the mean mass (\ref{mmeanI})
for $T \ge  0.5 \, T_H$ (or for $ \langle m \rangle \ge 0 $ ) to be equal to
the most probable mass of  bag from which one determines the resonance width:
\begin{eqnarray}\label{mmassII}
&  \hspace*{-.25cm}
\overline{m} (v)   \approx  \langle m \rangle  \quad {\rm and} \\
\label{mgammaII}
&  \hspace*{-.25cm}
\Gamma_R (v) \approx  2 \sqrt{2 \ln2 }\, \Gamma \hspace*{-.1cm} \left[
\frac{ \langle m \rangle  }{\textstyle B + \gamma^2 \beta}  \right]
\hspace*{-.1cm}  = 2 \,\gamma \sqrt{
\frac{2 \ln 2 \,\langle m \rangle }{B + \gamma^2 \beta}  }  \,.
\end{eqnarray}
The last two equations lead to a vanishing  ratio
$\frac{\Gamma_R}{\langle m \rangle} \sim  \langle m \rangle^{-\frac{1}{2}}$ in the limit $\langle m \rangle \rightarrow \infty$.
Comparing (\ref{mmassII}) and (\ref{mgammaII}) with the mass and width  (\ref{mgamI}) of the
Regge trajectory (\ref{alphaHT}) and applying absolutely the same logic which we used for the
free QGP bags, we conclude
that the location of the FWM heavy bags in the complex energy plane
is identical to that one of resonances belonging to the trajectory
(\ref{alphaHT}) with
\begin{eqnarray}\label{mgamII}
&
\langle m \rangle \approx |S|^\frac{1}{2} \quad {\rm and } \quad
a_r \approx  - 4 \gamma \, \sqrt{  \frac{\ln 2   }{  B + \gamma^2 \beta }  }
\,.
\end{eqnarray}
The  most remarkable output of such a conclusion  is that   the medium dependent FWM  mass and  width of the extended QGP bags obey the upper  bound  for  the Regge
trajectory asymptotic  behavior   obtained for point-like hadrons
\cite{Trushevsky:77}!

It is also  interesting that the resonance  width formula (\ref{mgammaII}) is generated by the most
probable volume
\begin{eqnarray}\label{vem}
v_E (m) \approx \frac{m}{\sqrt{B^2 + 2 \gamma^2 s^*} } = \frac{m}{B + \gamma^2 \beta}
\,.
\end{eqnarray}
of heavy resonances of mass $m \gg M_0$ that are described by the continuous  spectrum $F_Q (s,T)$   (\ref{FsHQ}).
This result can be  easily found by maximizing   the exponential in $F_Q (s,T)$
with respect to resonance volume $v$ at
fixed mass $m$  \cite{Reggeons:08}.

The extracted  values of the  resonance width coefficient along with the relation  (\ref{BI}) for  $B (T)$ allow us to estimate $a_r$  as
\begin{eqnarray}\label{aR}
&
a_r \approx  - 4  \, \sqrt{   \frac{2\,    T\, T_H \,  }{ 2\, T - T_H }   \ln 2 } \,.
\end{eqnarray}
This expression shows  that for $T \rightarrow T_H/2  + 0$  the asymptotic behavior
(\ref{alphaHT}) breaks down since the resonance width  diverges  at fixed $|S|$.
We hope that  such a behavior can be experimentally observed \cite{Bugaev:09} at  NICA (Dubna, Russia) and FAIR (Darmstadt, Germany) energies.

Now we can find the spin of the FWM  resonances
\begin{eqnarray}\label{Jm}
&
J  = {\rm Re} \, \alpha_r (\langle m \rangle^2) \, \, \approx   g_r^2  \,
 \langle m \rangle \left[ \langle m \rangle - \frac{a_r^2}{4} \right] \,,
\end{eqnarray}
which  has  a typical Regge behavior up to a small correction.
Such a property can also be  obtained   within  the dual resonance model \cite{DualRM},
within the models of  open  \cite{Shuryak:sQGP}  and  closed  string \cite{Shuryak:sQGP}, and within
 AdS/CFT \cite{AdS}.
These models  support our result (\ref{Jm}) and justify  it.  Note, however, that
in addition to the spin value  the FWM determines the width of hadronic  resonances.
The latter  allows us to predict   the  ratio
of widths of two resonances having spins $J_2$ and $J_1$
and appearing at the same temperature $T$  to be  as follows
\begin{eqnarray}\label{g2g1}
&
\frac{
\Gamma_R \biggl[ \frac{\langle m \rangle\bigl|_{J_2}}{ (B + \gamma^2 \beta) } \biggl]}{
\Gamma_R \biggl[ \frac{\langle m \rangle\bigl|_{J_1}}{ (B + \gamma^2 \beta)} \biggl]
}
\approx \frac{ \sqrt{v \bigl|_{J_2} }  }{
\sqrt{v \bigl|_{J_1}  } }
\approx \frac{ \sqrt{\langle m \rangle\bigl|_{J_2} }  }{
\sqrt{\langle m \rangle\bigl|_{J_1}  } } \approx   \left[  \frac{J_2}{J_1} \right]^\frac{1}{4}\,,
\end{eqnarray}
which, perhaps, can  be tested at LHC.

Now we turn to the analysis of the low temperature regime, i.e. to
$T \le  0.5 \,  T_H$.  Using previously obtained results from  (\ref{mmeanI}) one finds
\begin{eqnarray}\label{mmassIII}
&
\overline{m} (v)  ~ \approx ~  M_0 \,,
\end{eqnarray}
i.e.
the mean mass is volume independent. Taking the limit $v \rightarrow \infty$, we get the ratio $\frac{\Gamma(v)}{\overline{m} (v)} \rightarrow  \infty $ which closely resembles  the case of  the  lower bound  of the Regge trajectory asymptotics  (\ref{aLOW}).
Similarly to the analysis of  high temperature regime, from (\ref{aLOW}) one can find  the trajectory  phase and then  the resonance  mass $M_r$ and
its width $\Gamma_r$
\begin{eqnarray}\label{phiSL}
&
\phi_r  (S) \rightarrow - \pi + \frac{2 |C_l |  |\sin (\arg C_l) |  }{|S|^\frac{1}{2} }  \,,\\
\label{mgamL}
&
M_r \approx   |C_l |  |\sin (\arg C_l) | \quad {\rm and} \quad
\Gamma_r \approx  2 |S|^\frac{1}{2} \,.
\end{eqnarray}
Again comparing the  averaged  masses  and width of FWM resonances
with  their counterparts in  (\ref{mgamL}), we  find  similar behavior in the
limit of large  width of resonances.  Therefore, we
conclude that
at low temperatures the FWM obeys the lower bound of the Regge trajectory asymptotics of  \cite{Trushevsky:77}.

The other way of averaging, i.e. with respect to the resonance volume,
in the leading order   results in  an infinite value of the most probable resonance width \cite{Reggeons:08} defined in this way.
Note  that such a result is supported by  the high temperature mean width behavior  if   $T \rightarrow T_H/2 + 0$.
As one can see from (\ref{aR}) and (\ref{mgamI}),   in the latter case   the trajectory  (\ref{alphaHT})   also demonstrates
a very large width compared to a finite resonance  mass.

The FWM suggests that the Regge trajectories of QGP bags have a statistical nature.
The  above estimates  demonstrate  that at any temperature  the FWM QGP bags can be regarded as the medium induced Reggeons which at $T \le 0.5 \, T_H$
(i.e. for  $\langle m \rangle \le 0$) belong to the Regge trajectory   (\ref{aLOW}) and otherwise they are described by the  trajectory (\ref{alphaHT}).
Of course, both of the trajectories (\ref{aLOW}) and  (\ref{alphaHT})
are valid in the asymptotic $|S| \rightarrow \infty$, but the most remarkable  fact is that,
to our knowledge, the FWM gives us the first example of a model which reproduces both  of these trajectories and, thus, obeys both bounds of the Regge asymptotics.
Moreover, since the FWM contains the Hagedorn-like mass spectrum at any temperature,
it shows that  such a  spectrum is not exclusively  related to the linear Regge trajectories.  At low temperatures  (i.e. for  $\langle m \rangle \le 0$)  the large/heavy QGP bags  have the square root trajectory  (\ref{aLOW}) which is in line with expectations of
Ref. \cite{Trushevsky:77}.

\bigskip


\section{Conclusions}


Here we briefly describe  an exactly solvable statistical model, the FWM,  of the QGP equation of state.  It  accounts for the Hagedorn mass spectrum and for finite medium dependent width of large  QGP bags.  Inclusion of the Gaussian attenuation of the resonance  mass  leads  not only to a  convergency of partition function  at high temperatures, but also it allows us to explain a huge deficit of the experimentally  observed  hadronic resonances with masses above  $M_0 \approx 2.5$ GeV compared to the Hagedorn mass  spectrum.

The FWM allows us to establish the full  Regge trajectories
of  large/heavy QGP bags  both in a vacuum and in a strongly interacting medium.
The free QGP bags in a vacuum  have the linear Regge trajectory and, thus, they obey the
upper bound of the Regge trajectory asymptotics.  A linear Regge  trajectory  is also found for  the
in-medium QGP bags at temperatures above $0.5\, T_H$, whereas
for  temperatures  below $0.5\, T_H$   QGP bags   obey  the lower bound  of   the Regge trajectory asymptotics (the square root one) which is in line with   expectations of
Ref. \cite{Trushevsky:77}.

The FWM allows us to connect the statistical description of the QGP equation of state with
the Regge poles method and show that the Regge trajectories  of large/heavy QGP bags have a statistical nature.  These findings bring forward the statistical  models  of the  QGP equation of state to a  qualitatively new level of  realism.

\vskip3mm

{\bf Acknowledgments. }
The research made in this work
was supported in part   by the Program ``Fundamental Properties of Physical Systems
under Extreme Conditions''  of the Bureau of the Section of Physics and Astronomy  of
the National Academy of Science of Ukraine.
K.A.B. acknowledges  the partial support  by the Fundamental Research State Fund of Ukraine,
Agreement No F28/335-2009 for the Bilateral project FRSF (Ukraine) -- RFBR (Russia).

\end{document}